\documentclass[conference]{IEEEtran}

\usepackage{times}
\usepackage{epsfig}
\usepackage{amsmath}
\usepackage{amsfonts}
\usepackage{graphicx}
\usepackage{amssymb}
\usepackage{amstext}
\usepackage{latexsym}
\usepackage{color}
\usepackage{ifthen}
\usepackage{multirow}
\usepackage{verbatim}
\usepackage{array,tabularx}
\usepackage[mathscr]{euscript}
\usepackage{tikz}

\newcommand{\be}{\begin{equation}}
\newcommand{\ee}{\end{equation}}
\newcommand{\bear}{\begin{eqnarray}}
\newcommand{\eear}{\end{eqnarray}}
\newcommand{\bears}{\begin{eqnarray*}}
\newcommand{\eears}{\end{eqnarray*}}
\newcommand{\bi}{\begin{itemize}}
\newcommand{\ei}{\end{itemize}}
\newcommand{\ben}{\begin{enumerate}}
\newcommand{\een}{\end{enumerate}}

\newtheorem{theorem}{Theorem}
\newtheorem{lemma}[theorem]{Lemma}

\IEEEoverridecommandlockouts

\renewcommand{\baselinestretch}{1}

\begin{document}
\title{Data Secrecy in Distributed Storage Systems under Exact Repair}


\author{
\IEEEauthorblockN{Sreechakra Goparaju,  Salim El Rouayheb, Robert Calderbank and H. Vincent Poor}
\thanks{S. Goparaju, S. El Rouayheb, and H. Vincent Poor are with the Department of Electrical Engineering, Princeton University, USA (e-mails: {goparaju, salim, poor}@princeton.edu).}
\thanks{R.  Calderbank is with the Department of Computer Science, Duke University, USA (e-mail: robert.calderbank@duke.edu).}
\thanks{This research was supported by the U.\ S. National Science Foundation under Grant CCF-1016671.}

}
\maketitle
%

\begin{abstract}
The problem of securing data against eavesdropping in distributed storage systems is studied. 
The focus is on systems that use linear codes and implement exact repair to recover from node failures.
 The maximum file size that can be stored securely is determined for  
 systems in which all the available nodes help in repair (i.e.,  repair degree $d=n-1$, where $n$ is the total number of nodes) and for any number of compromised nodes. Similar results in the literature are restricted to the case of at most two compromised nodes. Moreover,  new explicit upper bounds are given on the maximum secure file size for systems with $d<n-1$. The key ingredients for the contribution of this paper are new results on subspace intersection for the data downloaded during repair. The new bounds imply the interesting fact that the maximum data that can be stored securely  decreases exponentially with the number of compromised nodes. 

 \end{abstract}


\section{Introduction}\label{sec:Intro}
We study the problem of making distributed storage systems (DSS) information-theoretically secure against eavesdropping attacks. These systems are witnessing a rapid growth in recent years and include data centers and p2p cloud storage systems. These systems use data redundancy to achieve data reliability and availability in the face of frequent node failures. Three-times (3x) data replication  has been the industry standard to achieve this goal. However, this solution does not scale well with the large amounts of data (in the order of petabytes) that these systems need to store. For this reason,  data centers have started utilizing more sophisticated erasure codes on part of their data (typically the ``cold" data that is not highly accessed) to protect against data loss \cite{LRCMicrosoft, XorbaFacebook}. 

\begin{figure}[t]
\begin{center}
\includegraphics[]{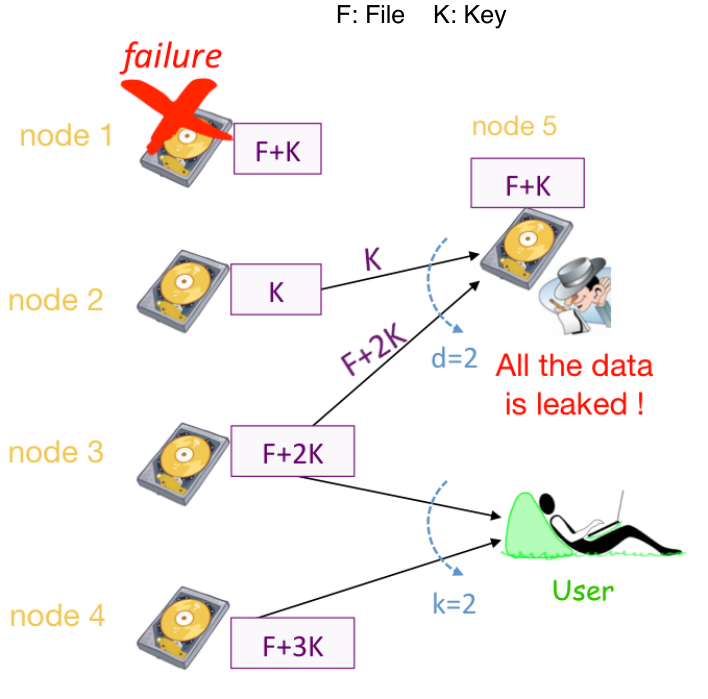}
\caption{ An example of how repairing a DSS can compromise the system security. The original DSS formed of nodes $1,\dots, 4$ is secured against a single compromised node using a secret sharing scheme or a coset  code. However, repairing failed nodes can break the security of the system. For instance, consider the case when  node 1 fails and is replaced by node 5, which is already compromised.  The eavesdropper can observe all the data downloaded by node 5 and therefore decode the stored file $F$. \vspace{-1cm}}
    \label{fig:ex1}
   \end{center}
\end{figure}
Erasure codes can achieve the same reliability levels as 3x replication with a much reduced storage overhead. However, they result in other system costs consisting of higher repair bandwidth, disk reads, computation complexity, etc. Moreover, erasure codes present new challenges when trying to secure the  system. We illustrate this phenomenon with the example in Fig.~\ref{fig:ex1}, which  depicts an $(n,k,d)=(4,2,2)$ DSS. The parameter $n=4$ represents the total number of nodes of unit storage capacity each, and  $k=2$ is the number of nodes contacted by a user to retrieve the stored file. A new node, added to the system after a failure, contacts $d=2$ other nodes to download its data ($d$ is referred to as the repair degree).  Fig.~\ref{fig:ex1} shows the failure and repair of node~$1$.   Using a maximum distance separable (MDS) code, such as a Reed-Solomon code, the user can store a file of size 2 units in the DSS, which is also the information-theoretically optimal size. Now, suppose that we want to protect the system against an eavesdropper that can observe at most one node in the DSS unknown to us. If the system does not experience failures and repairs, then one can store securely a file $F$ of one unit on the DSS by ``mixing" the information file $F$ with a randomly generated unit sized key $K$  using the code depicted in the figure. This code can be regarded as a secret sharing scheme \cite{Shamir}, a coset code for the wiretap channel II \cite{OW}, or as a secure network code for the combination network \cite{CY11, ES11}. The code allows a user contacting any 2 node to decode the file $F$ and leaks no information to the eavesdropper. A security violation occurs, however,  when a node fails and is replaced by a new one. The replacement node has to download data from the surviving nodes in the system to regenerate the lost data. Now, if the new node is already compromised, this will reveal all the downloaded data to the eavesdropper. For instance, the figure depicts the case when node 1 fails and the coded data chunk $F+K$ is lost. The new replacement node downloads the two data chunks $K$ and $F+2K$ to decode the lost packet $F+K$. However, this may reveal these two packets to the eavesdropper which can decode the file $F$. Therefore, even if we start with a perfectly secure code, the repair process can break the system security and result in data leakage. 
Our goal in this paper is to quantify how much data can be stored securely in a storage system even when the system experiences  failures and repairs.

We consider systems that implement \emph{exact repair}\footnote{ See \cite{DGWWR07} for the other type of repair referred to as \emph{functional} in the literature.} in which the repair process regenerates an exact copy of the lost packet (see Fig.~\ref{fig:ex1}). Exact repair is a requirement in many practical systems for numerous reasons, such as preserving the systematic form of the data and allowing temporary reconstruction of data stored on a ``hot" (i.e. highly accessed) node \cite{LRCMicrosoft}. We also focus on \emph{linear} coding schemes since they are the dominant class of codes employed in practice due to their ease of implementation.  For these systems we are interested in quantifying the maximum amount of data that can be stored in a DSS with a given storage and repair bandwidth budgets while keeping the system \emph{perfectly secure}. This means that we want to guarantee that no  information is leaked to an eavesdropper that can observe a certain number of nodes in the system. 

\noindent\paragraph*{Contribution} We find an expression for the maximum file size that can be stored securely on a DSS under the linear coding and exact repair constraints. Our result holds for any number of compromised nodes for  a DSS with repair degree $d=n-1$. Similar results  in the literature exist  only for systems in which at most two nodes can be compromised by an eavesdropper.
 We also give new explicit upper bounds on the maximum secure file size for systems with $d<n-1$. The key ingredients for our contribution are new results on subspace intersection for the data downloaded during repair.  Our bounds imply the interesting fact that the maximum secure file size  in the minimum storage regime decreases exponentially with the number of compromised nodes in contrast with for example the minimum-bandwidth regime \cite{PRK11, RSKGlobecom11} or secret sharing schemes where it decreases linearly. 

%
%
%
%
%

\noindent\paragraph*{Related work} Dimakis et al. studied in \cite{DGWWR07} the information-theoretic tradeoff between storage overhead and repair bandwidth in distributed storage systems. Pawar et al. studied the problem of securing distributed storage systems under repair dynamics against eavesdroppers and malicious adversaries in \cite{PRK11,PRK10, PRKISIT11}. They provided upper bounds on the system secure capacity and proved its achiveability in the bandwidth-limited regime for repair degree $d=n-1$. Shah et al. constructed  secure codes  based on the product-matrix framework in \cite{RSKGlobecom11} and \cite{RSK12}. These codes can achieve the upper bound in \cite{PRK10} for the minimum-bandwidth regime and for any repair degree $d$.  Rawat et al. gave tighter bounds on the secrecy capacity of a DSS in the minimum storage regime \cite{RKSV13} and proved the achieveability of their bound for $d=n-1$ and for certain system parameters. Dikaliotis et al. studied the security of distributed storage systems in the presence of a trusted verifier \cite{DDH10}.

\noindent\paragraph*{Organization} The paper is organized as follows. In Section~\ref{sec:Model}, we describe the system and eavesdropper models and set up the notation. In Section~\ref{sec:results}, we state our main results. We follow these by first providing an intuition behind the results in Section \ref{sec:intuition} and then the proofs in Section~\ref{sec:proofs}.  We conclude with a summary of our results and open problems in Section~\ref{sec:conclusion}.


\section{Problem Setting}\label{sec:Model}

\subsection{System Model}

A distributed storage system  consists of $n$ \emph{active} storage units or nodes $\{1, 2, \ldots, n\}$, each with a storage capacity of $\alpha$ symbols belonging to some finite field $\mathbb{F}$. Nodes in a DSS are unreliable and fail frequently. When a storage node fails, it is  replaced by a new node with the same storage capacity $\alpha$.  A DSS  storing a data file ${\cal F}$ of $M$ symbols (in $\mathbb{F}$) allows any legitimate user called a data collector to  retrieve the $M$ symbols and reconstruct the original file ${\cal F}$ by connecting to any $k$ out of the $n$ active nodes. We term this the \emph{MDS property} of the DSS. Furthermore, we focus on single  node failures since they are the most frequent in such systems. A new node added to the system to replace a failed one connects to  $d$ arbitrary nodes  chosen out of the remaining $n-1$ active ones and downloads $\beta$ units from each. 
The repair degree $d$ is a system parameter satisfying $k \le d \le n-1$, and the nodes aiding in the repair are called \emph{helper} nodes. The so-called \emph{repair} process usually demands a higher repair bandwidth $d \beta$ than the amount of data $\alpha$ it actually stores. Moreover, the reconstructed data can possibly be different from the original data stored in the failed node. We define an $(n,k,d)$-DSS as a DSS that uses $d$ nodes for the repair of a failed node to continuously maintain the $k$-out-of-$n$ MDS property.

Dimakis et al. \cite{DGWWR07} showed that there is a fundamental tradeoff between the amount of data stored in each node $\alpha$ and the minimum repair bandwidth $d \beta$ required to store a file in the system. We focus on one extremity of this tradeoff, called minimum storage, in which each node stores the minimum possible $\alpha = M/k$. An MDS code  achieving the minimum repair bandwidth for this $\alpha$,
\begin{eqnarray}\label{eq:orb}
d \beta &=& \frac{d\alpha}{d-k+1},
\end{eqnarray}
is referred to as an optimal bandwidth MDS code or a minimum storage regenerating (MSR) code. Furthermore, in this paper, we consider the case of \emph{exact repair}, where the replacement node is required to reconstruct an exact copy of the lost data. In other words, the DSS consisting of $n$ active nodes (and the MSR code) is invariant with time.  It has been shown that optimal repair bandwidth is achievable for exact repair \cite{DRWS10}. 

We concentrate on the practical scenario of linear MSR codes, which preserve the optimal repair bandwidth of (\ref{eq:orb}). Without loss of generality, we can separate the nodes in the DSS storing an MDS code into systematic and parity nodes. We designate the first $k$ nodes as systematic, where node $i, i \in [k] := \{1, 2, \ldots, k\}$, stores the data vector $w_i$ of column-length $\alpha$. The data vector $w_{k+i}$ stored in parity node $i, i \in [n-k]$, is given by
\begin{eqnarray}\label{eq:paritynodes}
w_{k+i} &=& \sum_{j=1}^k A_{i,j}w_j,
\end{eqnarray}
where $A_{i,j} \in\mathbb{F}^{\alpha \times \alpha}$ is the \emph{coding matrix} corresponding to the parity node $i \in [n-k]$ and the systematic node $j \in [k]$. For optimal bandwidth repair of a failed systematic node $i \in [k]$, all other nodes transmit $\beta$ amount of information, i.e., a helper node $j \neq i$ transmits a vector of length $\beta$ given by $V_{j,i}w_j$, where $V_{j,i} \in \mathbb{F}^{\beta \times \alpha}$ is the repair matrix used for the repair of node $i$ by node $j$. 
The vector $V_{j,i}w_j$ can also be interpreted as a projection of $w_j$ onto a subspace of dimension $\beta$. We will use $V_{j,i}$, interchangeably, to denote both the matrix and the subspace obtained by the span of its rows.

\subsection{Eavesdropper Model}
We assume the presence of an eavesdropper Eve in the DSS, which can passively observe but not modify the contents of up to $\ell < k$ nodes of its choice.  
Eve can not only observe the data stored in a node $i$, but also the repair data  $V_{j,i}w_j$  flowing into its replacement from a helper node $j \neq i$. In other words, not only does Eve have complete knowledge of $w_i$, it can potentially infer a part of $w_j$ as well. In line with our assumption of repair of only systematic nodes, we assume that Eve can observe the repair data for only a subset of the systematic nodes\footnote{Note that for securing the data, we do not store the original file on the systematic nodes, but rather the original file data encoded with random keys. However, we shall continue to refer to these nodes as systematic for convenience.}, ${\cal E}_d$, where ${\cal E}_d \subseteq [k]$, and denote the rest of the observed nodes (for which it just observes the stored data) as ${\cal E}_s$, ${\cal E}_s \subseteq [n]$. The size of these subsets are denoted by $\ell_1 = |{\cal E}_s|$ and $\ell_2 = |{\cal E}_d|$, where $\ell_1 + \ell_2 = \ell$.
Finally, we assume that Eve has
complete knowledge of the storage and repair schemes implemented in
the DSS.
 

\subsection{Secrecy Capacity}
Let $U$ be a random vector uniformly distributed over
$\mathbb{F}^{M^{(s)}}$, representing an incompressible data file with
$H(U) = M^{(s)}$.
Let $W_i$ denote the random variable corresponding to the data $w_i$ stored in node $i$, $i \in [n]$. 
Let us assume that a set ${\cal D}$ of $d$ helper nodes aid in the repair of node $i$.
We denote the random variable corresponding to the data transmitted by a helper node $m \in {\cal D}$ for the repair of node $i$ by $S_m^i({\cal D})$, and the total repair data downloaded by node $i$ by $S^i_{{\cal D}}$. We drop the ${\cal D}$ in the notation and call these $S_m^i$ and $S^i$, respectively, when the context is clear.

Thus, $W_i$
represents the data that can be downloaded by a data collector when
contacting node $i$ and observable by Eve when $i \in {\cal E}_s$, while $S^i$
represents the total data revealed to Eve when accessing a node
$i \in {\cal E}_d$. Notice that the stored data $W_i$ is a function of the downloaded data $S^i$. For convenience let us denote $\{W_i: i \in {\cal A}\}$ by $W_{\cal A}$, $\{S_i^j: j \in {\cal A}\}$ by $S_i^{\cal A}$, and $\{S^i: i \in {\cal A}\}$ by $S^{\cal A}$.

The MDS property of the DSS can be written as
\begin{eqnarray}\label{eq:rp}
H\left(U\left|W_{\cal A}\right.\right) &=& 0,
\end{eqnarray}
for all ${\cal A} \subseteq [n]$, such that $|{\cal A}| = k$. To store a file $U$ on the DSS perfectly secured from the eavesdropper Eve, we have the \emph{perfect secrecy} condition,
\begin{eqnarray}\label{eq:ps}
H\left(U\left|W_{{\cal E}_s}, S^{{\cal E}_d}\right.\right) &=& H\left(U\right),
\end{eqnarray}
for all ${\cal E}_s \subseteq [n], {\cal E}_d \subseteq [k] \backslash {\cal E}_s$, and $|{\cal E}_s| +|{\cal E}_d| < k$.


Given an $(n,k,d)$-DSS with $\ell_1$ and $\ell_2$ compromised nodes (as described above) its linear coding secrecy capacity $C_s(\alpha)$, is then defined to be the maximum file size $H(U)$ that can be stored in the DSS using an optimal bandwidth MDS code for exact repair, such that the reconstruction property and the perfect secrecy condition simultaneously hold, i.e.,
\begin{eqnarray}\label{eq:secrecy_capacity}
C_s (\alpha):= \sup_{\substack{{\cal A}, {\cal E}_s, {\cal E}_d: \\(\ref{eq:rp}), (\ref{eq:ps}) \,{\rm hold}} }H(U).
 \end{eqnarray}



%
%




\section{Main Results}\label{sec:results}
In this section, we state our main results. The proofs will follow in Section \ref{sec:proofs} after give a rough idea behind the results in Section \ref{sec:intuition}. The following lemma provides a lower bound on the sum of subspaces\footnote{The sum of subspaces $B$, $C$ is defined as $B+C = \{b+c:b\in B, c\in C\}$.} associated with the repair bandwidth from a particular node, when it aids in the repair of multiple nodes.
\begin{lemma}\label{lem1}
Consider an $(n,k,d)$-DSS in the systematic form with nodes having storage capacity $\alpha$. Let nodes $[k]$ be the systematic nodes and let $V_{i,j}$ be the $\beta \times \alpha$ matrix associated with the exact repair of node $j$ by node $i$. Then, for $d = n-1$ and for each $i \in [k]$, we have
\begin{eqnarray}\label{eq:lem1}
\dim\left(\sum_{j \in {\cal A}}V_{i,j}\right) &\ge& \left(1 - \left(\frac{n-k-1}{n-k}\right)^{|{\cal A}|}\right)\alpha,
\end{eqnarray}
where ${\cal A} \subseteq [k]\backslash \{i\}$ and $V_{i,j}$ is the subspace corresponding to the matrix.
\end{lemma}

\smallskip
The next theorem gives an upper bound on the (linear coding) secrecy capacity $C_s(\alpha)$ for a given number of compromised nodes.
\begin{theorem}\label{thm1}
Consider an $(n,k,d)$-DSS with a node storage capacity of $\alpha$, which stores an optimal bandwidth linear MDS code for exact repair of systematic nodes. Suppose an eavesdropper gains access to the data stored in $\ell_1$ nodes and the data downloaded during the repair of $\ell_2$ systematic nodes, such that
\begin{eqnarray*}
\ell_1 + \ell_2 &<& k.
\end{eqnarray*}
The achievable secure file size $M^s$ for the given MSR code is then upper bounded by
\begin{eqnarray}\label{eq:thm1}
M^{s} &\le& (k-\ell_1-\ell_2)\left(1-\frac{1}{d-k+1}\right)^{\ell_2}\alpha.
\end{eqnarray}
\end{theorem}

\smallskip
The next theorem establishes that the upper bound in Theorem \ref{thm1} is achievable when $d = n-1$.
\begin{theorem}\label{thm2}
For an $(n, k, d)$-DSS with $d = n-1$, the secrecy capacity for optimal bandwidth MDS codes such that any systematic node is exact-repairable using a linear coding, is achievable for $\alpha = (n-k)^k$ and is given by
\begin{eqnarray}
C_s(\alpha) &=& (k-\ell_1-\ell_2)\left(1-\frac{1}{n-k}\right)^{\ell_2}\alpha.
\end{eqnarray}
Moreover, the capacity is achievable for all $(\ell_1, \ell_2)$.
\end{theorem}
\begin{IEEEproof}
The proof follows from \cite[Theorem 10]{RKSV13} which describes an achievability scheme by precoding a ($n,k$) zigzag code \cite{TWB13} using a maximum rank distance code.
\end{IEEEproof}

\section{Some Intuition}\label{sec:intuition}
Before giving the formal proofs, we present in this section some intuition behind the upper bound (\ref{eq:thm1}) on the  maximum achievable secure file size for a given DSS. We start with a simple  toy example. 
How much data can we store securely in a system of $2$ nodes of storage size $\alpha$ such that a user can recover it in the presence of an eavesdropper which has access to any one (unknown to us) node? We can quickly upper bound the answer by $\alpha$ data units by arguing that if we actually knew which node was compromised, we would not store any information in that node. In fact, this upper bound is achievable by using a random key $r$ of size $\alpha$ and storing it on node $1$ and storing  $w + r$ on node $2$, where $w$ is the data. In other words, we subtract the amount of information visible to the eavesdropper from the total storage size available to the user. Then, by exploiting randomness this upper bound can be achieve even without knowing the  identity of the compromised nodes.

\begin{figure}
\centering
{
\begin{tikzpicture}[scale=0.4,>=stealth]
\draw[blue] (0,2) circle (1cm);
\fill[blue] (0,6) circle (0.15cm);
\fill[blue] (0,5) circle (0.15cm);
\fill[blue] (0,4) circle (0.15cm);
\draw[blue] (0,8) circle (1cm);
\draw[blue] (0,11) circle (1cm);
\draw[blue] (0,14) circle (1cm);

\draw[red] (10,4) circle (1cm);
\fill[red] (10,8) circle (0.15cm);
\fill[red] (10,7) circle (0.15cm);
\fill[red] (10,6) circle (0.15cm);
\draw[red] (10,10) circle (1cm);
\draw[red] (10,13) circle (1cm);

\draw[->, gray] (1,2) -- (9,4); \draw[->, gray] (1,2) -- (9,10); \draw[->, gray] (1,2) -- (9,13);
\draw[->, gray] (1,8) -- (9,4); \draw[->, gray] (1,8) -- (9,10); \draw[->, gray] (1,8) -- (9,13);
\draw[->, gray] (1,11) -- (9,4); \draw[->, gray] (1,11) -- (9,10); \draw[->, gray] (1,11) -- (9,13);
\draw[->, gray] (1,14) -- (9,4); \draw[->, gray] (1,14) -- (9,10); \draw[->, gray] (1,14) -- (9,13);

\draw (-6,2) node[right] {node $k$}; \draw (-6,8) node[right] {node $\ell+3$}; \draw (-6,11) node[right] {node $\ell+2$}; \draw (-6,14) node[right] {node $\ell+1$};
\draw (12,4) node[right] {node $\ell$}; \draw (12,10) node[right] {node $2$}; \draw (12,13) node[right] {node $1$};
\draw (5,13.5) node[above] {$V_{\ell+1,1}w_{\ell+1}$};
\draw (5,3) node[below] {$V_{k,\ell}w_{k}$};
\end{tikzpicture}
}
\caption{ An $(n,k,n-1)$-DSS  in which  nodes $1,\dots,\ell$ have failed and have been replaced by the $\ell$ compromised nodes in red (nodes on the right). Each red node is repaired by contacting all of the other $d=n-1$ nodes in the system. For clarity, we only depict the edges between the red nodes and the remaining $k-\ell$ non-compromised systematic nodes (in blue, on the left).
At a high level, the upper bound in Theorem \ref{thm1} is obtained by evaluating the amount of information  leaked to the eavesdropper in this scenario. This includes the information stored in the red nodes, plus the information contained in their downloaded data. The latter can be  bounded using Lemma~\ref{lem1} which gives a handle on the correlation among all the data downloaded for repair.\vspace{-.5cm} }
\label{fig:intuition}
\end{figure}
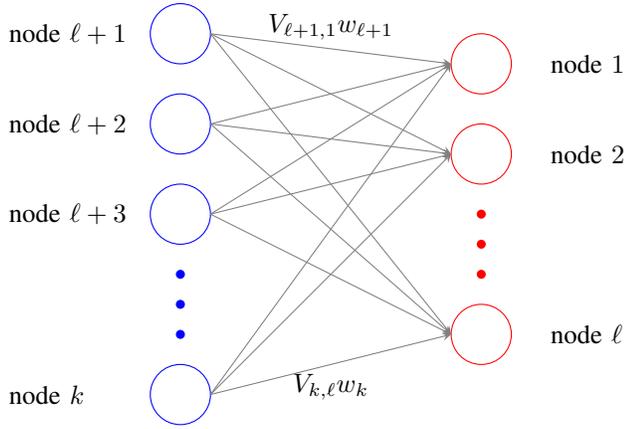


We extend this argument to an $(n,k,d)$-DSS with nodes of storage capacity $\alpha$ and repair bandwidth $\beta$ per helper node. We explain our results for the specific case of two parity nodes and $d = n-1$ for which the optimal repair bandwidth  $\beta = \alpha/2$. This means that each helper node sends half of its ``information'' to a replacement node.
We also restrict our attention to the more-compromised nodes for which Eve can observe the repair data, i.e., let $\ell = \ell_2 = |{\cal E}_d|$, where ${\cal E}_d \subseteq [k]$. As in the $2$-node example, we first try to find an upper bound on the maximum secure file size by asking how much would we store if we knew that the first $\ell$ nodes were compromised, or ${\cal E}_d = [\ell]$. We do not store any information on these $\ell$ nodes. We know however that Eve also gains some information about the nodes aiding in the repair of the compromised nodes if they were to fail. We assume that in a large enough length of time, each node fails at least once, and is repaired by the rest of the nodes. In particular, Eve has access to the information flowing to each of the compromised nodes from each of the remaining nodes. Fig. \ref{fig:intuition} shows the information flows which we shall focus on.

The information observable by Eve about node $i$, $i \in \{\ell+1, \ldots, k\}$, is obtained by the vectors $V_{i,j}w_i$ communicated from node $i$ to each compromised node $j \in [\ell]$. The total knowledge Eve has about node $i$ is therefore equivalent to the combined information present in $\{V_{i,j}w_i\}_{j=1}^{\ell}$, or in other words, equivalent to the rank of the $\beta \times \ell \alpha$ matrix,
\begin{eqnarray*}
\big[V_{i,1}\left|\right.V_{i,2}\left|\right.\cdots \left|\right.V_{i,\ell}\big].
\end{eqnarray*}
If instead, we view $V_{i,j}$ to be a subspace spanning the rows of the matrix $V_{i,j}$, this rank can also be represented as the dimension of the sum of subspaces,
\begin{eqnarray*}
\dim\left(V_{i,1} + V_{i,2} + \cdots + V_{i,\ell}\right).
\end{eqnarray*}

In this paper, we provide explicit bounds for the dimension of these sums of subspaces. For two parity nodes, we show that an addition of each repair subspace from a node reveals half of the information (about the helper node) which was unrevealed before the addition. To clarify, all subspaces reveal half of the information by design ($\beta = \alpha/2$). If we add two subspaces, because any two of these subspaces, say $V_{i,1}$ and $V_{i,2}$, cannot intersect in more than $\alpha/4$ dimensions \cite{TWB12, RKSV13}, their sum has to be more than
\begin{eqnarray*}
\frac{\alpha}{2} + \frac{\alpha}{4} 
\end{eqnarray*}
dimensions. Lemma~\ref{lem1} implies  that for $\ell$ subspaces, a lower bound of
\begin{eqnarray*}
\frac{\alpha}{2} + \frac{\alpha}{4} + \cdots + \frac{\alpha}{2^{\ell}} &=& \left(1 - \frac{1}{2^{\ell}}\right)\alpha
\end{eqnarray*}
dimensions has to be revealed by node $i$ in repairing the $\ell$ compromised nodes.

This calculation thus gives us the amount of information visible to Eve, which is $\ell \alpha$ from the compromised nodes and
\begin{eqnarray*}
(k-\ell)\left(1 - \frac{1}{2^{\ell}}\right)\alpha
\end{eqnarray*}
from the $(k-\ell)$ non-compromised nodes. As in the $2$-node example, it can be proved that using randomness (maximum rank separable codes, \cite{RKSV13}), we can securely store a total of $k\alpha$ minus the information visible to Eve, i.e., 
\begin{eqnarray*}
(k-\ell)\,\frac{1}{2^{\ell}}\,\alpha
\end{eqnarray*}
data units in the presence of $\ell$ compromised nodes.


\section{Proofs}\label{sec:proofs}
\begin{IEEEproof}[Proof of Lemma \ref{lem1}]
We prove the lemma for the case of two parity nodes, i.e., $n = k+2$. The proof can easily be extended to the case of more than two parity nodes. For the corresponding $(k,k+2,d=k+1)$-DSS, as in Section \ref{sec:Model}, we represent the symbols stored in the nodes $[n]$ by the column-vectors $w_1, \ldots, w_n$ of length $\alpha$, and assume the first $k$ nodes to be systematic. For convenience, we rename the coding matrices for parity node 1, $A_{1,j}, j \in [k]$ as $A_j, j \in [k]$, and those for parity node 2, $A_{2,j}, j \in [k]$ as $B_j, j \in [k]$.


When node $j$ fails, node $i$ transmits the matrix $V_{i,j}w_i$ in order to repair node $j$. When the number of parity nodes is $2$, $V_{i,j}$ is an $\alpha/2 \times \alpha$ matrix. For notational simplicity, we represent the matrices $V_{k+1,j}$ and $V_{k+2,j}$ by $S_{1,j}$ and $S_{2,j}$ for all $j \in [k]$. It can be shown that an optimal bandwidth exact repair of systematic nodes necessitates interference alignment \cite{SRKR12-4} and leads to the following \emph{subspace conditions} (e.g. \cite{TWB12}):
\begin{eqnarray}\label{eq:sc1}
S_{1,j}A_i &\backsimeq&  S_{2,j}B_i,\\
\label{eq:sc2}
&\backsimeq& V_{i,j},\\
\label{eq:sc3}
S_{1,j}A_j + S_{2,j}B_j &\backsimeq& \mathbb{F}^{\alpha},
\end{eqnarray}
for all $j \in [k], i \in [k]\backslash \{j\}$, and $\backsimeq$ denotes an equality of subspaces. In other words, the above subspace equalities specify the conditions required for the repair of a systematic node $j$ by the set of helper nodes $[n]\backslash \{j\}$.

We prove the result stated in the lemma using induction.
\emph{Base case:} For $|{\cal A}| = 1$, we have
$\dim\left(V_{i,j}\right) \ge \alpha/2$,
which follows from the model constraints on the given DSS\footnote{It can be shown from the MDS property of the storage code that all the coding matrices $\{A_i, B_i\}, i\in [k]$ have full rank, and from the subspace conditions that all the subspaces $V_{i,j}$ have full rank as well.}.

\smallskip
\emph{Inductive step:} Suppose the claim holds for $|{\cal A}| = m-1$.
We shall prove that the claim also holds for $|{\cal A}| = m$.
Without loss of generality, let ${\cal A} = [m]$.

For $[k] \ni i \not \in [m]$, we have
\begin{eqnarray}\label{eq:proof1}
\dim\left(\sum_{j=1}^{m}V_{i,j}\right) &=& \dim\left(\sum_{j=1}^{m}S_{1,j}A_i\right),\\
\label{eq:proof2}
&=& \dim\left(\sum_{j=1}^{m}S_{1,j}\right),\\
\label{eq:proof3}
&=& \dim\left(\sum_{j=1}^{m}S_{1,j}A_{m}\right),\\
&\ge& \dim\left(\sum_{j=1}^{m-1}S_{1,j}A_{m} \cap S_{2,m}B_{m}\right) \nonumber\\
\label{eq:proof4}
&&+ \dim\left(S_{1,m}A_{m}\right),
\end{eqnarray}
where (\ref{eq:proof1}) follows from (\ref{eq:sc2}), 
(\ref{eq:proof2}) and (\ref{eq:proof3}) follow 
from distributivity and
the fact that the matrices $A_i$ and $A_{m}$ are invertible, and therefore $\dim\left(SA_i\right) = \dim\left(S\right) = \dim\left(SA_{m}\right)$, for any subspace $S$. For (\ref{eq:proof4}), notice that the subspaces
\begin{eqnarray*}
\left(\sum_{j=1}^{m-1}S_{1,j}A_{m}\right) \cap S_{2,m}B_{m},
\end{eqnarray*}
and
$S_{1,m}A_{m}$ intersect only in the zero vector, see (\ref{eq:sc3}). Furthermore, both are contained in the subspace
\begin{eqnarray*}
\left(\sum_{j=1}^{m}S_{1,j}A_{m}\right),
\end{eqnarray*}
and hence so is their \emph{direct} sum.

Using the identity for arbitrary subspaces $S_a$ and $S_b$, that
$\dim\left(S_a + S_b\right) + \dim\left(S_a \cap S_b\right) = \dim\left(S_a\right) + \dim\left(S_b\right)$,
and the fact that the subspaces $S_{2,m}B_m$ and $S_{1,m}A_m$ have dimension $\alpha/2$ ($A_m$ and $B_m$ being nonsingular), we obtain from (\ref{eq:proof4}),
\begin{eqnarray}
\label{eq:proof5}
\dim\left(\sum_{j=1}^{m}V_{i,j}\right) &\ge& \dim\left(\sum_{j=1}^{m-1}S_{1,j}A_m\right) + \alpha\\
&&- \dim\left(\sum_{j=1}^{m-1}S_{1,j}A_{m} + S_{2,m}B_{m}\right)\nonumber.
\end{eqnarray}

The third term on the right hand side in inequality (\ref{eq:proof5}) equals the term on the left hand side, because
\begin{eqnarray}
\lefteqn{\dim\left(\sum_{j=1}^{m-1}S_{1,j}A_{m} + S_{2,m}B_{m}\right)}\nonumber\\
&=& \dim\left(\sum_{j=1}^{m-1}S_{1,j}A_{m}B_{m}^{-1} + S_{2,m}\right)\nonumber\\
&=& \dim\left(\sum_{j=1}^{m-1}S_{2,j} + S_{2,m}\right)\nonumber\\
\label{eq:proof6}
&=& \dim\left(\sum_{j=1}^m V_{i,j}\right),
\end{eqnarray}
where the steps follow from similar reasons as in (\ref{eq:proof1})--(\ref{eq:proof4}). Also, similarly,
\begin{eqnarray}
\label{eq:proof7}
\dim\left(\sum_{j=1}^{m-1}S_{1,j}A_m\right) &=& \dim\left(\sum_{j=1}^{m-1}V_{m,j}\right).
\end{eqnarray}

Using the induction hypothesis and (\ref{eq:proof5})--(\ref{eq:proof7}), we have
\begin{eqnarray}
2\dim\left(\sum_{j=1}^m V_{i,j}\right) &\ge& \dim\left(\sum_{j=1}^{m-1}V_{m,j}\right) + \alpha\nonumber\\
&\ge& \left(1 - \frac{1}{2^{m-1}}\right)\alpha + \alpha,\nonumber
\end{eqnarray}
which completes the inductive step.
\end{IEEEproof}

\bigskip
For the sake of completeness, we present here the information-theoretic proof given in \cite{RKSV13} which transitions into the proof of Theorem \ref{thm1} via Lemma \ref{lem1}. However, our notation, described in Section \ref{sec:Model}, is inspired by \cite{SRKR12}. 
\begin{IEEEproof}[Proof of Theorem \ref{thm1}]
Let ${\cal R}$ be any set of $k-\ell_1-\ell_2$ systematic nodes not in ${\cal E}_s$ or ${\cal E}_d$. In order to store a file $U$ of entropy $M^{(s)}$ securely in the DSS, we have
\begin{eqnarray}
\label{eq:pf1}
M^{(s)} &=& H\left(U \left| W_{{\cal E}_s}, S^{{\cal E}_d}\right.\right),\\
\label{eq:pf2}
&=& H\left(U \left| W_{{\cal E}_s}, S^{{\cal E}_d}\right.\right) - H\left(U \left| W_{{\cal E}_s}, S^{{\cal E}_d}, W_{{\cal R}}\right.\right), \IEEEeqnarraynumspace \\
&=& I\left(U; W_{{\cal R}}\left| W_{{\cal E}_s}, S^{{\cal E}_d}\right.\right),\\
&\le&  H\left(W_{{\cal R}}\left| S^{{\cal E}_d}\right.\right),\\
&\le& \sum_{i \in {\cal R}} H\left(W_i \left| S_i^{{\cal E}_d}\right.\right),\\
&=& \sum_{i \in {\cal R}} \left(H\left(W_i, S_i^{{\cal E}_d}\right) - H\left(S_i^{{\cal E}_d}\right)\right),\\
\label{eq:pf3}
&=& \sum_{i \in {\cal R}} \left(H\left(W_i\right) - H\left(S_i^{{\cal E}_d}\right)\right),
\end{eqnarray}
where (\ref{eq:pf1}) is the same as (\ref{eq:ps}), (\ref{eq:pf2}) follows from (\ref{eq:rp}) and the fact that $W^i$ is a function of $S^i$, and (\ref{eq:pf3}) from the fact that $S_i^m$ is a function of $W_i$, for any $m \neq i$.
Using the linearity of the MDS code being used, we have
\begin{eqnarray}
H\left(S_i^{{\cal E}_d}\right) &=& \dim\left(\sum_{j \in {\cal E}_d}V_{i,j}\right),
\end{eqnarray}
where because $i \in {\cal R}$, we have ${\cal E}_2 \subseteq [k]\backslash \{i\}$. Thus, we have,
\begin{eqnarray}
M^{(s)} &\le& \left(k-\ell_1- \ell_2\right)\left(1 - \frac{1}{n-k}\right)^{\ell_2}\alpha.
\end{eqnarray}

For $d < n-1$, we focus on the first $d+1$ nodes, viewing it as an $(n' = d+1, k, d = n'-1)$-DSS. The conditions of exact repair for this restricted system form a relaxation of the original problem, and thus an upper bound on $M^{(s)}$ for this system also holds for the latter. By the optimal bandwidth condition (\ref{eq:orb}),
and because exact repair requires interference alignment, from Lemma \ref{lem1}, we obtain for $i \in [k]$,
\begin{eqnarray*}
\dim\left(\sum_{j \in {\cal A}}V_{i,j}\right) &\ge& \left(1-\left(\frac{d-k}{d-k+1}\right)^{|{\cal A}|}\right)\alpha,
\end{eqnarray*}
for ${\cal A} \subseteq [k] \backslash \{i\}$. Note that our helper set of nodes is ${\cal D} = [n']\backslash \{i\}$ for repairing node $i$. We therefore obtain the required bound on $M^{(s)}$ using a similar set of equations as for $d = n-1$.
\end{IEEEproof}

\section{Conclusion}\label{sec:conclusion}

We have studied the problem of securing data in distributed storage systems against eavesdropping. Our focus has been on systems that implement linear codes and exact repair. We have determined the maximum file size that can be stored securely in these systems for any number of compromised nodes, when the repair degree $d=n-1$.  For the other cases, \emph{i.e.}, when $d<n-1$, we have given new upper bounds on the amount of secure data that can be stored in the system. Many questions remain open, such as  constructing codes that can achieve our upper bound \eqref{eq:thm1} for $d<n-1$, and finding   a general expression of the system secrecy capacity  without the linearity and exactness assumptions.

\end{document}